\documentclass[apjl]{emulateapj}
\usepackage{color}

\usepackage{graphicx}

\begin{document}

\title{First Images of the Protoplanetary Disk around PDS 201}

\color{black}
\shorttitle{Images of the Disk around PDS 201}
\shortauthors{Wagner et al.}
\author{Kevin Wagner$^{1,2,\star}$, Jordan Stone$^{1,3}$, Ruobing Dong$^{4}$, Steve Ertel$^{1,5}$, Daniel Apai$^{1,2,6}$, David Doelman$^{7}$, Alexander Bohn$^{7}$, Joan Najita$^{2,8}$, Sean Brittain$^{9}$, Matthew Kenworthy$^{7}$, Miriam Keppler$^{10}$, Ryan Webster$^{1}$, Emily Mailhot$^{1}$, \& Frans Snik$^{7}$}


\altaffiltext{1}{Steward Observatory, University of Arizona}
\altaffiltext{2}{NASA NExSS \textit{Earths in Other Solar Systems} Team}
\altaffiltext{3}{NASA Hubble Postdoctoral Fellow}
\altaffiltext{4}{University of Victoria, British Columbia, Canada}
\altaffiltext{5}{Large Binocular Telescope Observatory}
\altaffiltext{6}{Lunar and Planetary Laboratory, University of Arizona}
\altaffiltext{7}{Leiden Observatory, Leiden University, The Netherlands}
\altaffiltext{8}{National Science Foundation Optical/IR Laboratory}
\altaffiltext{9}{Clemson University, South Carolina}
\altaffiltext{10}{Max Planck Institute for Astronomy, Heidelberg, Germany}

\altaffiltext{$\star$}{Correspondence to: kwagner@as.arizona.edu}

\keywords{Stars: pre-main sequence (PDS 201)  --- protoplanetary disks  --- planet-disk interactions}

\begin{abstract}

Scattered light imaging has revealed nearly a dozen circumstellar disks around young Herbig Ae/Be stars$-$enabling studies of structures in the upper disk layers as potential signs of on-going planet formation. We present the first images of the disk around the variable Herbig Ae star PDS 201 (V* V351 Ori), and an analysis of the images and spectral energy distribution through 3D Monte-Carlo radiative transfer simulations and forward modelling. The disk is detected in three datasets with LBTI/LMIRCam at the LBT, including direct observations in the $Ks$ and $L'$ filters, and an $L'$ observation with the 360$^\circ$ vector apodizing phase plate coronagraph. The scattered light disk extends to a very large radius of $\sim$250 au, which places it among the largest of such disks. Exterior to the disk, we establish detection limits on substellar companions down to $\sim$5 M$_{Jup}$ at $\gtrsim$1$\farcs$5 ($\gtrsim$500 au), assuming the \cite{Baraffe2015} models. The images show a radial gap extending to $\sim$0$\farcs$4 ($\sim$140 au at a distance of 340 pc) that is also evident in the spectral energy distribution. The large gap is a possible signpost of multiple high-mass giant planets at orbital distances ($\sim$60-100 au) that are unusually massive and widely-separated compared to those of planet populations previously inferred from protoplanetary disk substructures. 


\end{abstract}

\section{Introduction}

Recent observational advances have ushered in a new era of high-angular resolution studies of protoplanetary disks, and have revealed that substructures are a common$-$if not ubiquitous$-$property (e.g., \citealt{Muto2012}, \citealt{Andrews2018}, \citealt{Huang2018}). This has led to the prevailing hypothesis that on-going planet formation is common among these systems. Such structures that are frequently hypothesized to be linked to forming planets include gaps, rings, spiral arms, and vortices, among other structures.

Within this context, disks around Herbig Ae/Be stars (Herbig disks) represent a key component in our understanding of the formation of planetary systems. Wide-orbit ($\gtrsim$10 au) massive ($\gtrsim$3-5 M$_{Jup}$) planets are more frequent around higher-mass stars (\citealt{Nielsen2019}, \citealt{Wagner2019b}), suggesting that Herbig disks may be the best places to look for ongoing giant planet formation and signs of planet-disk interactions. Recent images of Herbig Ae/Be and lower-mass T Tauri systems have shown gaps cleared by forming planets (\citealt{Keppler2018}, \citealt{Wagner2018a}), spiral arms driven by stellar companions (\citealt{Dong2016}, \citealt{Wagner2018b}), and in some cases, spirals possibly generated by planetary-mass companions (\citealt{Wagner2019}). Meanwhile, radio interferometric observations trace midplane (i.e. density) features that are not directly accessible in scattered light (e.g., \citealt{Andrews2018}), and have revealed the presence of gas kinematic features (e.g., \citealt{Teague2019}) that are possibly associated with planets interior to disk gaps as well as planets possibly driving spiral arms (\citealt{Pinte2020}).

A major challenge in understanding the statistical properties of Herbig disks remains the overall low sample size. Of the ten Herbig disks imaged around single stars prior to this study, two show prominent two-armed spirals, while nearly all show signatures of gaps and rings (\citealt{Dong2018}). These occurrence rates, which potentially trace massive planets forming on wide orbits, are much higher than detection rates of 1-10\% for wide-orbit giant planets\footnote{Again, planets with masses $\gtrsim$3-5 M$_{Jup}$ and semimajor axes $\gtrsim$10 au.} around slightly older stars (e.g., \citealt{Bowler2016}, \citealt{Stone2018}, \citealt{Nielsen2019}) assuming high initial entropy planetary evolution models (e.g., \citealt{Baraffe2015}). While this result could simply reflect the selection biases of Herbig disk surveys, the high spiral fraction could hint at an abundant population of wide-orbit giant planets that have eluded detection to date.

Since wide-orbit giant planets are thought to be the primary driver of two-armed spirals (e.g., \citealt{Dong2015b}), differences in the occurrence rates of two-armed spirals and the occurrence rates of giant planets themselves may indicate that a population of giant planets exists that is less luminous than typical model assumptions (e.g., \citealt{Marley2007}), and/or could point to important details about the migratory timescales of wide-orbit giant planets. Because imaging surveys of Herbig disks to date are biased toward objects previously surveyed or otherwise known to possess interesting features, imaging surveys of unbiased samples of Herbig disks are needed to explore the above ideas.

Here, we take a step in this direction in reporting the first scattered light images of the Herbig disk around PDS 201, which we selected as a Herbig disk target with no reported disk images. PDS 201 (V* V351 Ori) is a Herbig A7Ve member of the Orion OB1 association (\citealt{Ripepi2003}, \citealt{Hernandez2005}, \citealt{Alecian2013}). Our primary motivation was to image the substructures within the disk, and to thereby increase the number of imaged Herbig disks for statistical studies of disk substructures. The basic properties of the star are listed in Table 1. The star has most notably been studied for its variability (e.g., \citealt{vandenAncker1996}, \citealt{Ripepi2003}). However, unlike more typical variable Herbig Ae stars such as UX Ori, PDS 201 has also shown contrasting periods of quiescence. This behavior is possibly associated with episodic accretion (\citealt{vandenAncker1996}). Depending on its orientation, extinction from the inner disk may also play a role in the observed stellar variability. Determining the disk orientation and its contribution to the star's optical variability is a secondary objective of our observations.


\begin{deluxetable}{lcc}
\tabletypesize{\scriptsize}
\tablecaption{Properties of PDS 201}
\tablewidth{0pt}
\tablehead{\colhead{Parameter} & \colhead{Value} & \colhead{Ref.}}

\startdata
Spectral Type & A7V & 1\\
Mass & $\sim$2.0 M$_\odot$ &  1 \\
T$_{eff}$ & 7400 K & 1\\
Age & 1$-$6.5 Myr & 1,2\\
Distance & 342 $\pm$5 pc & 3\\
$Ks$ & 6.8 mag & 4\\
$L^\prime$ & 5.8 mag & 5
\enddata
\tablecomments{(1) \citealt{Ripepi2003}, (2) \citealt{vandenAncker1996}, (3) \citealt{GAIADR2}, (4) \citealt{Cutri2003}, (5) \citealt{WISE}.}
\end{deluxetable}

\section{Observations and Data Reduction}

We observed PDS 201 in three different imaging modes with the Large Binocular Telescope (LBT) Interferometer (LBTI). The basic properties of the observations and data reduction can be found in Table 2. Each observation utilized the LBT $L-$ and $M$-band Infrared Camera (LMIRCam, \citealt{Leisenring2012}) located behind the cryogenic beam combiner of the LBTI (\citealt{Hinz2016}).  The LBTI typically combines the light from the two 8.4m apertures; however, due to on-going upgrades only one aperture was in operation. Each observation consisted of a several hours long imaging sequence with a substantial amount of field rotation (60$-$80$^\circ$) and periodic telescope nods. For the first two observations, no coronagraph was used (direct imaging) and the core of the primary star was saturated on the detector. For these observations, photometric calibration sequences with shorter on-chip exposure times (0.2$-$0.5 sec) were taken at the beginning and end of the observation. 


For the third dataset, we utilized the recently installed double-grating vector apodizing phase plate (vAPP) coronagraph (\citealt{Doelman2017}), which was designed to improve exoplanet and disk imaging capabilities at small angular separations. PDS 201 is the first disk observed with the new optic. The vAPP suppresses the Airy pattern of all sources in the focal plane, reducing their intensity by orders of magnitude between 2.7 and 15 $\lambda$/D. Furthermore, the vAPP is a pupil-plane coronagraph and thus the coronagraphic performance is unaffected by nodding or telescope vibrations. However, suppressing the Airy rings comes at the cost of reducing the Airy core throughput by a factor of 2.2. This reduction of throughput of the central star and off-axis sources alike enables the central PSF core to be used as a photometric calibrator for the full observing sequence.

\begin{deluxetable}{lccc}
\tabletypesize{\scriptsize}
\tablecaption{Observing Log}
\tablewidth{0pt}
\tablehead{\colhead{Parameter} & \colhead{19-Nov-2019} & \colhead{5-Jan-2020} & \colhead{7-Jan-2020}  }

\startdata
Filter & $L^{\prime}$  & $Ks$  & $L^{\prime}$ \\
Coronagraph & Direct  & Direct  & vAPP \\
Avg. Seeing  & 0$\farcs$7  & 1$\farcs$0 & 0$\farcs$7 \\
Field Rot.  & 76$^\circ$ & 82$^\circ$ & 67$^\circ$ \\
Exp. Time & 0.91s &1.97s & 0.81s\\
Total Int. Time  & 3.5 hr & 2.5 hr & 3.0 hr\\
\hline
\smallskip\\
\multicolumn{4}{c}{Data Reduction Parameters} \\
\hline\\
Frame Binning & 100 &  50 & 100\\
KLIP Components & 7 & 7 & 10 \\
Ann. Segments & 8 & 8 & 8 \\
Radial Range & 8-150 px & 8-150 px & 8-150 px\\ 
Ref. Angle Range& 0.3-76$^\circ$  & 0.5-82$^\circ$  & 0.5-67$^\circ$
\enddata
\end{deluxetable}

After subtracting the first read of each detector ramp from the last to remove reset noise (correlated double sampling) and replacing bad pixels in the data, we subtracted the background via a running median of the nearest 250 sky frames. We then aligned the images via cross-correlation, and determined the precise image center via rotational-based centering \citep{Morzinski2015}. We identified and removed bad frames as those with a maximum cross-correlation of less than 0.95 with respect to the median image, which resulted in $\sim$10\% frame rejection. We binned the frames and performed PSF modelling and subtraction via Karhunen-Lo\`eve Image Projection (KLIP: \citealt{Soummer2011}) using the implementation in \cite{Apai2016}. Finally, we derotated the images and combined them using using a noise weighted mean \citep{Bottom2017}.

\section{Results}




\subsection{Disk Structures}

\begin{figure*}[htpb]
\figurenum{1}
\epsscale{1.18}
\plotone{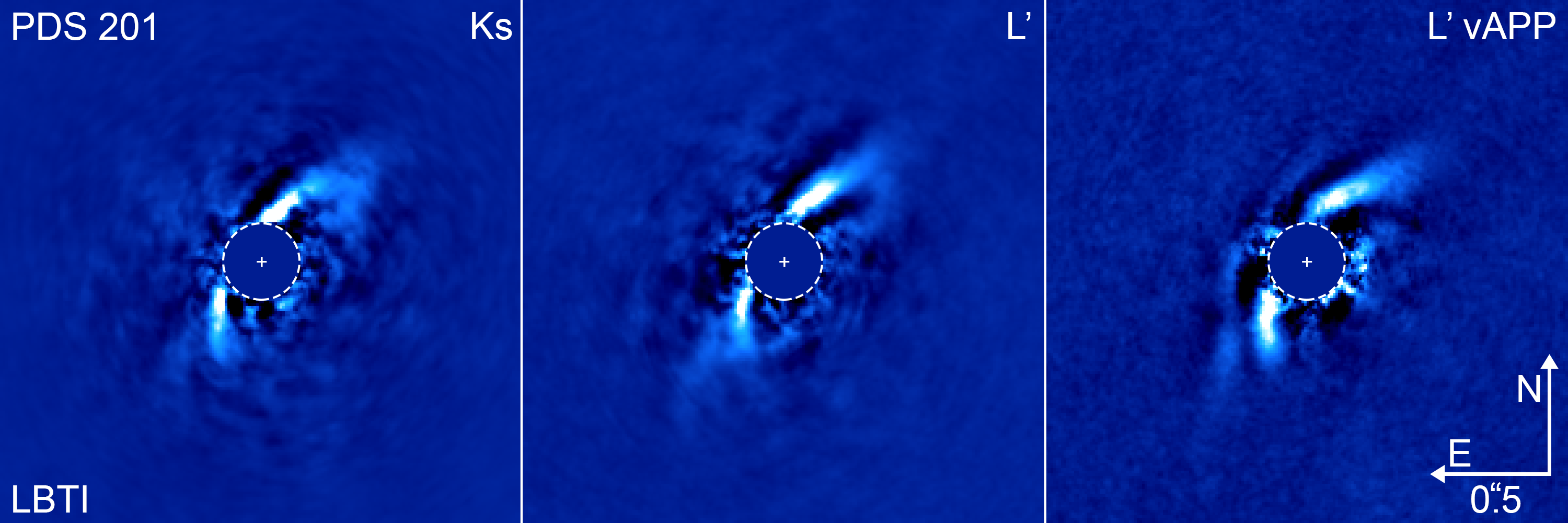}
\caption{From left to right: direct $Ks$ image of PDS 201; $L^\prime$ direct image, and $L^\prime$ + vector apodized phase plate (vAPP) image. Each image clearly shows the upper forward-scattering disk surface extending to $\sim$0$\farcs$8 along the major axis, while the image taken with the vAPP shows what is likely the forward-scattering surface of the far-side of the disk (see illustrations in Figure 2 for more details).}
\end{figure*}

\begin{figure*}[htpb]
\figurenum{2}
\epsscale{1.18}
\plotone{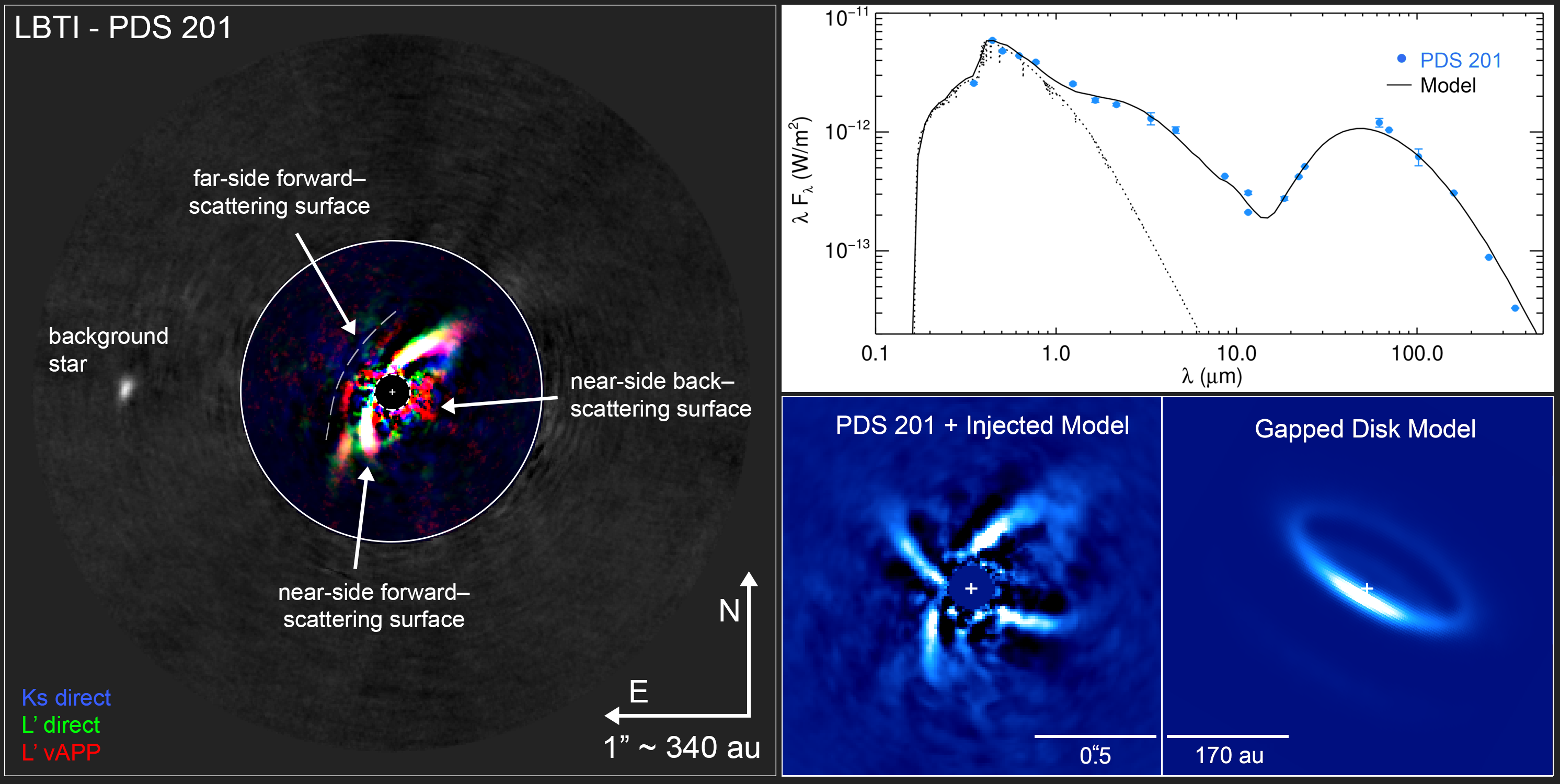}
\caption{Left: Schematic diagram of the PDS 201 system. All images show the prominent trace of the disk's near-side scattering surface to the Northeast, while the vAPP image (red) also reveals the far-side forward-scattering surface further to the Northeast. Each image also show a very low-signal to noise detection of the back-scattering surface of the near-side of the disk to the Southwest. Top right: Spectral energy distribution of PDS 201 and model disk. Bottom Right: processed and original $Ks$ model disk images.}
\end{figure*}

The images are shown in Figure 1. The disk is clearly detected in each image, primarily as the bright forward-scattering surface from the upper layers of the near-side of the disk (see diagram in Figure 2). By fitting an ellipse to the bright forward-scattering surface, we estimate a disk inclination of $\sim$70$^\circ \pm 10^\circ$ and position angle of $\sim -$35$^\circ \pm 5^\circ$ E of N. The gapped nature of the disk is evident from the double-peaked spectral energy distribution (SED; top-right panel of Figure 2), which shows a deficit of flux at $\sim$10$-$20 $\mu$m and a second peak at $\sim$50 $\mu$m that is consistent with a gap size of $\sim$140 au for a central A7-type star. The association of such an SED profile with gapped disks has been well-established in Herbig disk studies. For example, the SED of MWC 758 has a similar dip at shorter wavelengths, corresponding to a $\sim$70 au gap \citep{Grady2013}. The vAPP image also revealed an arc of scattered light adjacent to the disk to the Northeast, which is likely the forward-scattering surface of the far-side of the disk. Likewise, all three observations show low signal-to-noise arc-like emission to the Southwest, which is likely the back-scattering surface of the near-side of the disk. 

We utilized the \texttt{HOCHUNK3D} Monte-Carlo radiative transfer simulation software \citep{Whitney2013} to construct a model of the disk to compare to the images and SED. Our primary aim was to constrain the bulk disk properties (i.e., the disk size and location of the gap) without biases introduced by the data reduction. To compare to the processed images of the disk, we injected the model into the $Ks$ data prior to running the KLIP algorithm at an orientation roughly perpendicular to the PDS 201 disk.\footnote{The $Ks$ data was chosen as it has the best angular resolution.} The original and processed model images are shown in the bottom-right panels of Figure 2, while the photometric data and model parameters are tabulated in the appendix.

We found a good match to the images and SED with a 250 au disk including a gap from 1$-$140 au at an inclination of 65$^{\circ}$. The size of the disk and outer radius of the gap are the two best-constrained parameters, while the inclination is degenerate with the flaring exponent. This model is meant to be a simple representation, and we note that while the primary disk geometry is reasonably well-constrained, there are a range of grain compositions, flaring exponents, etc. that provide an equivalent fit to the data. A more detailed model may also utilize the location of the far-side forward scattering surface revealed by the vAPP data to more precisely constrain the disk scale height, inclination, and flaring, although we defer this possibility to future work.


\subsection{Limits on Companions Exterior to the Disk}

The $Ks$ and direct $L^\prime$ images show a source at 1$\farcs$7 to the East. The faintness ($L^\prime \sim$17.7, $Ks \sim$18.2) of the source combined with its relatively neutral $Ks-L^\prime$ color is indicative of a background star. No other planet candidates were identified. To determine the image sensitivity, we performed extensive source injection and recovery tests. We injected simulated sources in the data (prior to running the KLIP algorithm) along ten equally spaced position angles beginning with 0$^\circ$ PA and from separations of 0$\farcs$2 to 2$\farcs$5 in radial steps of 0$\farcs$1. We reduced the brightness of the injected sources and repeated each reduction until the source was recovered with a signal to noise ratio of $\sim$5, as calculated via Equation 9 of \cite{Mawet2014}, while also excluding any apertures contaminated by scattered light from the bright near-side forward-scattering surface or the background star at 1$\farcs$7. 

The resulting azimuthally-averaged contrast curves are shown in Figure 3. At small radii, few apertures are available, and most are at some level contaminated by disk signals, which vary between datasets. Therefore, the contrast in the inner regions is likely underestimated and may be affected differently by the amount of disk signals in each individual dataset. Even so, Figure 3 shows that the vAPP offers an improved contrast between 450 mas and 800 mas by up to a factor of two. Beyond 800 mas the vAPP reaches the background limit, which is $\sim$5.5 times higher for the vAPP observation. This can be explained by the reduction of core throughput by a factor of 2.2, which results in a relative increase of background photon noise by a factor of 4.8, and by the 14\% shorter exposure time. 

\begin{figure*}[htpb]
\figurenum{3}
\epsscale{1.18}
\plotone{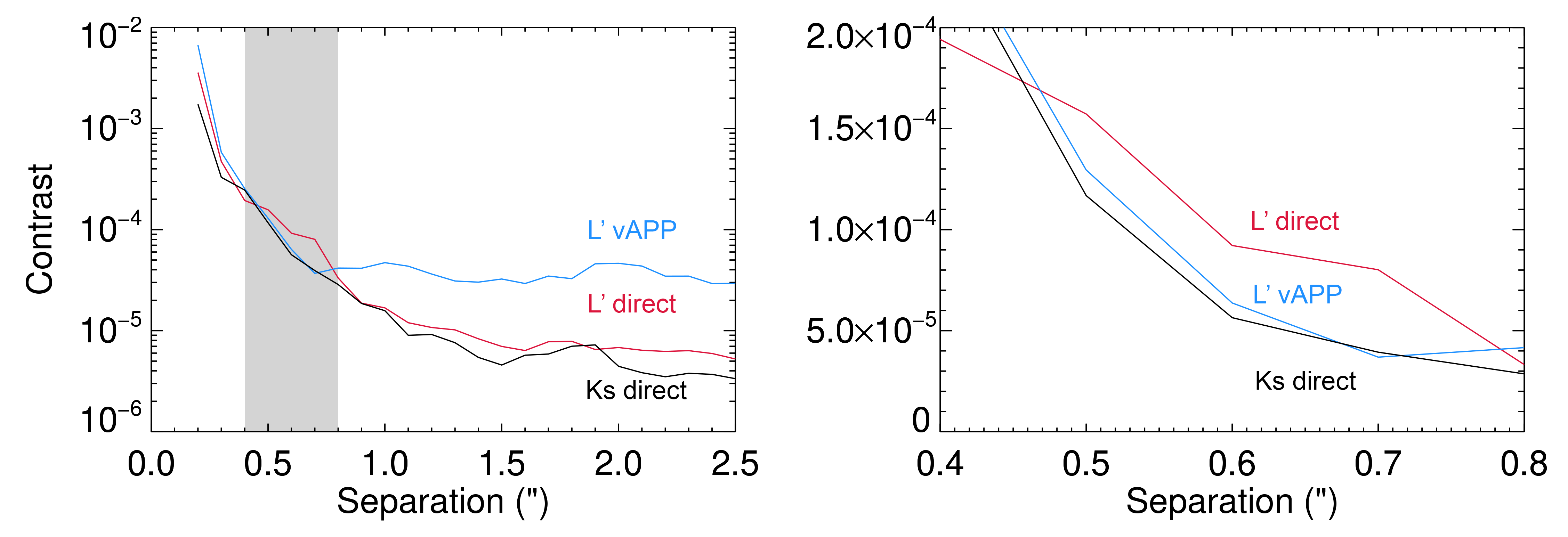}
\caption{Left: 5$\sigma$ contrast sensitivity vs. separation for the LBTI observations of PDS 201. The gray-shaded region corresponds to the angular range where the vAPP coronagraph has better contrast performance (and therefore shows a cleaner morphology of the disk), which is shown on the right in a zoomed-in version. Aside from the disk and a background star located at 1$\farcs$7 from the central star, no source is detected above 5 sigma in multiple epochs.}
\end{figure*}

Due to the disk's relatively high inclination, the images are only potentially sensitive to companions interior to the disk at specific orbital phases. Thus, we report only detection limits on companions exterior to the disk, for which the direct $L^\prime$ data provides the deepest limits in terms of mass. We utilize the hot-start models of \cite{Baraffe2015} for conversion between absolute magnitude to mass and assume an age of 1-10 Myr, which is consistent with the range of age estimates in the literature (\citealt{vandenAncker1996}, \citealt{Ripepi2003}). 

Along the observed minor axis of the disk, companions could be seen at $\sim$0$\farcs$5 with contrasts of $\sim$5$\times$10$^{-4}$, corresponding to a mass $\gtrsim$30 M$_{Jup}$. At $\sim$1$\farcs$0, companions could be seen along any axis with contrasts of $\sim$2$\times$10$^{-5}$, corresponding to a mass $\gtrsim$7 M$_{Jup}$, and at $\gtrsim$1$\farcs$5, companions could be seen with contrasts of $\sim$1$\times$10$^{-5}$, corresponding to a mass $\gtrsim$5 M$_{Jup}$. Utilizing the cold-start models instead (e.g., \citealt{Fortney2008}) results in higher mass detection limits that extend to approximately the deuterium burning limit, although such planets may be less common \citep{Marleau2019}.

\section{Discussion}

\subsection{Substructures in PDS 201}

Our primary motivation to observe PDS 201 was to improve upon the statistics of scattered light substructures that are potentially related to planet formation$-$namely gaps and spirals. To this effect, the detection of the disk around PDS 201 raises the number of imaged disks around (apparently) single Herbig Ae/Be stars from ten to eleven, as compared to the previous analysis of \cite{Dong2018}. Like most Herbig disks, PDS 201 has a wide gap extending to $\sim$0$\farcs$4 along the major-axis, or $\sim$140 au at the system's distance of 340 pc \citep{GAIADR2}. For a reasonable model comparison of a 70$^\circ$ inclined disk with a gap opened by a several M$_{Jup}$ planet, see also Fig. 15 of \cite{Dong2016c}. We also note a small asymmetry in the SE vs. NW sides of the disk$-$specifically, the SE side of the bright ring appears slightly shorter in projected separation (by $\sim$100 mas), and the fainter far-side exhibits a apparent kink at the SE-most extent that could possibly trace the flaring of the disk. 

There is no obvious indication of spiral arms in the disk around PDS 201. While spirals could be hidden by the relatively high inclination of the disk, interior to the gap, or simply because they are too faint to detect, we proceed under the assumption that PDS 201 lacks spiral arms.\footnote{This assumption does not qualitatively change our conclusions.} In this case, the total number of imaged Herbig disks around single stars with (observed) two-armed spirals is unchanged by the addition of PDS 201, thereby slightly lowering the occurrence rate of such spirals from 20$^{+13}_{-8}$\% to 18$^{+11}_{-7}$\%. These numbers remain uncertain due to the volume-limited nature of the sample, but despite the low number of observations the occurrence rate of two-armed spirals (those thought to be driven by massive planets: e.g., \citealt{Dong2015b}, \citealt{Bae2018b}) appears to be higher than typical estimated occurrence rates for wide-orbit giant planets of ($\sim$1-10\%: e.g., \citealt{Bowler2016}, \citealt{Stone2018}, \citealt{Nielsen2019}). While the observed occurrence rate of two-armed spirals among this sample is marginally consistent with an intrinsic occurrence rate of $\sim$10\%, there is less than a 5\% chance that the intrinsic occurrence rate is 3\%, and a 0.5\% chance that the intrinsic rate is 1\%.



The relatively high occurrence rate of two-armed spirals could perhaps indicate one or more of the following: 1) that two-armed spirals can be launched by less-massive and/or colder planets than predicted by current models, 2) that giant planets migrate inward on timescales of $\sim$10 Myr to orbits at which they are undetectable to direct imaging, or 3) that the two-armed spirals do not reliably indicate the presence of giant planets. The first scenario yields a testable prediction, as such a population of wide-orbit low-mass and/or cold-start giant planets (down to $\lesssim$1 M$_{Jup}$ at a few tens of au) could be easily identified by the upcoming \textit{James Webb Space Telescope} (JWST: e.g., \citealt{Beichman2019}) or upcoming 30-m-class telescopes. Meanwhile, the second option is seemingly unlikely given the long migration timescales that result from low disk gas masses at ages of $\gtrsim$10 Myr and low disk densities at $\gtrsim$50 au. The third possibility would require an alternative mechanism to explain their origin. Potential alternatives so far have invoked gravitational instabilities, which would require unrealistically high disk masses (e.g., \citealt{Kratter2016}) or spirals caused by shadowing-effects, which would require chance synchronizations of the outer disk orbit with misaligned inner disk precession timescales (\citealt{Montesinos2016}). 

\subsection{Comparison to Other Large Disks}

The PDS 201 disk has a size of at least $\sim$250 au, which places it at the upper end in the disk size distribution measured in near-infrared scattered light. In a recent survey, \citet{avenhaus18} and \citet{garufi20} imaged 29 stars in polarized light at $H$-band using VLT/SPHERE. Four sources (IM Lup, RXJ 1615, DoAr 25, and V1094 Sco) have a well-defined disk extending to 250 au or larger (two more sources, WW Cha and J1615-1921, appear to show structures at $r\gtrsim250$ au, however it is unclear those are part of a coherent disk). Similarly, the Strategic Explorations of Exoplanets and Disks with Subaru Survey \citep[SEEDS]{tamura09} imaged 68 young stellar objects (including other Herbig Ae/Be targets) at $H$-band using Subaru/HiCIAO \citep{uyama17}, and found only a handful with a circumstellar disk larger than PDS 201's (e.g., AB Aur; \citealt{hashimoto11}). Note that many disks have sizes measured in mm continuum dust emission; however, these observations trace the distribution of $\sim$mm-sized dust. The comparison of the sizes of these disks with those seen in scattered light$-$especially those tracing sub-$\micron$-sized dust$-$is not straightforward, as different constituents may have different spatial distribution due to mechanisms such as dust-gas interactions \citep{weidenschilling77}.


The large size of the central cavity in the disk of PDS201 is also unusual. Earlier studies have shown that multiple multi-Jupiter mass planets are needed to clear such large extended regions of a protoplanetary disk (\citealt{Zhu2012}; \citealt{Dods2011}). The planets detected in the disk of PDS 70 fit these expectations: the two planets have masses of 2-17 M$_{Jup}$ located at orbital distances $\sim$40\% and $\sim$70\% of the cavity radius (\citealt{Muller2018}, \citealt{Haffert2019}). If similar planets are responsible for the 140 au central cavity of PDS201, they would reside at $\sim$60 au and $\sim$100 au, and their masses would likely be higher than the PDS70 planets due to the larger stellar mass of PDS201 ($\sim$2 M$_{\odot}$) compared to PDS70 (0.8 M$_{\odot}$; \citealt{Muller2018}). 

This range in orbital distance (60-100 au) and planet mass (5-20 M$_{Jup}$), is unusual compared to the planet populations previously inferred from protoplanetary disk structures. Analyses of the disk gaps found in the DSHARP survey (\citealt{Zhang2018}), the Taurus sample of \cite{Long2018}, and individual disks collected by \cite{Bae2018} have inferred the presence of many low mass planets ($<$ 2 M$_{Jup}$) at orbital distances 60-100 au, but have not inferred the presence of any higher mass planets (\citealt{Lodato2019}). In contrast, direct imaging searches for planets around older (post-Herbig) stars have identified companions in this range of mass and radius, e.g., Kappa And b (\citealt{Carson2013}; $\sim$13-20 M$_{Jup}$ at 100 au) and HR8799b (\citealt{Marois2008}; $\sim$5-7 M$_{Jup}$ at 70 au). PDS 201 may be an evolutionary precursor of such systems. 


The large orbital distances of these planets is a challenge to traditional planet formation theories that rely on the ballistic collision of planetesimals to grow critical mass planetary cores capable of accreting large amounts of gas before the disk dissipates (\citealt{Pollack1996}). The main hurdle is forming a core fast enough given the extended dynamical timescales at large stellocentric radius and disk lifetimes of only a few million years. If forming planets between 60 and 100 au are responsible for the gap seen in PDS 201,  then their existence gives some support to more modern theories that can accelerate the formation of planetessimals and their subsequent growth to planetary cores (e.g., \citealt{Youdin2005}, \citealt{Ormel2010}). Further detailed study of PDS201, including searches for orbiting planets within its disk cavity, could lend new insights into how such planetary systems form.

\subsection{Disk Contribution to Variability}

PDS 201 exhibits $V$-band variability with a $\sim$0.6 mag amplitude over timescales of days to months \citep{ASASSN} and also displays a variable emission line profile (e.g., \citealt{YSOVAR}, \citealt{vandenAncker1996}, \citealt{Ripepi2003}). The fact that PDS 201 hosts a protoplanetary disk and also exhibits variability of the central star are likely related to some degree$-$for instance, an edge-on disk could contribute significantly to the broadband optical variability. Indeed, the \textit{outer} disk is seen to be highly inclined at $\sim$70$^{o}$. Given the long orbital timescales, this component likely does not play a significant role in the variability. However, if the inner disk is nearly co-planar with the outer disk, then variable absorption within this component would likely be a significant source of optical variability (e.g., \citealt{Bouvier2013}). 

\subsection{Prospects for Future Observations}

The first images of the disk around PDS 201 also raise new and exciting possibilities. Notably, the presence of a large gap in the disk makes the system a good target for searches for forming planets. Recently, two accreting protoplanets were found around a similar young star, PDS 70 (\citealt{Keppler2018}, \citealt{Wagner2018a}, \citealt{Haffert2019}). The geometry of the disk around PDS 201 appears to be similar to that of the disk around PDS 70, and its more extended nature could be indicative of a more massive and more extended planetary system.

Additionally, while our observations revealed the basic geometry of the disk in scattered light, radio interferometry with ALMA can probe the disk's gas kinematics and midplane dust distribution. These observations will more precisely pinpoint the locations of dust gaps and rings, and could also reveal kinematic tracers of forming planets (e.g., \citealt{Teague2019}, \citealt{Pinte2020}). Finally, differential polarimetry (e.g., \citealt{deBoer2020}) can potentially outperform the total intensity scattered light images presented here.

\section{Summary and Conclusions}

1. We presented the first images of the disk around PDS 201 (V* 351 Ori), taken with the Large Binocular Telescope. Remarkably, PDS 201's disk is one of the largest seen, extending to $\sim$250 au. The iamges show a disk gap extending to $\sim$140 au, while the disk shows no obvious large-scale spiral structures.

2. We modelled the scattered light disk and spectral energy distribution (SED) through 3D Monte-Carlo radiative transfer simulations and forward-modelling. This analysis confirmed that the apparent $\sim$140 au gap is consistent with both the second peak in the SED and the structures observed in the images.

3. We combined multiple near-infrared filters ($Ks$ and $L^\prime$) to characterize both the disk and a point source located at 1$\farcs$7 to the East. The point source's photometry suggests it is likely a background star.

4. We computed mass-detection limits from synthetic planet injections exterior to the disk, and find that planets down to $\sim$5 M$_{Jup}$ can be excluded at large separations ($\gtrsim$1$\farcs$5) from the star, assuming the \cite{Baraffe2015} models.

5. We incorporated the observed properties of the disk around PDS 201 into the statistics of disk substructures seen around Herbig stars in scattered light, which slightly lowers the occurrence rate of disks around single Herbig stars showing prominent two-armed spirals to 18$^{+11}_{-7}$\%. This is higher than typical occurrence rates of wide-orbit giant exoplanets, which, along with \cite{Dong2018}, we speculate may indicate the presence of a larger population of low-mass and/or cold-start planets that could be detected in the future with JWST. 

6. We suggest that PDS 201, as a recently discovered Herbig disk, offers significant opportunities for follow-up studies to reveal the inner disk structure and potential planetary system residing within the gap of the remarkably large disk around PDS 201. 

\section{Acknowledgments} The authors express their sincere gratitude to LBT Director Christian Veillet for allocating director's time for this project and to Sebastiaan Haffert for helping to plan the vAPP observation. The results reported herein benefited from collaborations and/or information exchange within NASA's Nexus for Exoplanet System Science (NExSS) research coordination network sponsored by NASA's Science Mission Directorate. J.M.S. is supported by NASA through Hubble Fellowship grant HST-HF2-51398.001-A awarded by the Space Telescope Science Institute, which is operated by the Association of Universities for Research in Astronomy, Inc., for NASA, under contract NAS5-26555. The research leading to these results has received funding from the European Research Council under ERC Starting Grant agreement 678194 (FALCONER). The LBT is an international collaboration among institutions in the United States, Italy, and Germany. LBT Corporation partners are: The University of Arizona on behalf of the Arizona university system; Istituto Nazionale di Astrofisica, Italy; LBT Beteiligungsgesellschaft, Germany, representing the Max-Planck Society, the Astrophysical Institute Potsdam, and Heidelberg University; The Ohio State University, and The Research Corporation, on behalf of The University of Notre Dame, University of Minnesota, and University of Virginia.


\appendix

In this appendix, we provide additional data tables that were used for our analysis but were not essential to the discussion at hand. For the SED analysis in \S3.1, we assembled data from the literature spanning wavelengths of 0.35 $\mu$m to 350 $\mu$m, whice we provide in Table 3. The SED shows a clear double-peaked form indicative of a transition disk with large gap (e.g., \citealt{Strom1989}). In Table 4, we provide the parameters of the model disk presented in \S3.1. 

\clearpage

\begin{deluxetable}{cccc}
\tabletypesize{\scriptsize}
\tablecaption{Archival PDS 201 Photometry}
\tablehead{\colhead{$\lambda$ ($\mu$m)} & \colhead{Flux (W/m$^2$)} & \colhead{Uncertainty (W/m$^2$)} & \colhead{Ref.}}

\startdata

0.350 &  2.57$\times 10^{-12}$ & 5.09$\times 10^{-14}$ &  1\\
0.444 &  5.90$\times 10^{-12}$ & 1.08$\times 10^{-13}$ & 2\\
0.505 &  4.80$\times 10^{-12}$ & 1.01$\times 10^{-13}$ & 3\\
0.623 &  4.39$\times 10^{-12}$ & 2.41$\times 10^{-14}$ & 3\\
0.772 &  3.87$\times 10^{-12}$ & 6.60$\times 10^{-14}$ &  3 \\
1.24 &   2.54$\times 10^{-12}$ & 4.84$\times 10^{-14}$ & 4\\
1.65 &   1.85$\times 10^{-12}$ & 7.27$\times 10^{-14}$ &  4\\
2.16 &   1.63$\times 10^{-12}$ & 4.11$\times 10^{-14}$ &  4 \\
3.35 &   1.30$\times 10^{-12}$ & 1.52$\times 10^{-13}$ & 5\\
4.60 &   1.04$\times 10^{-12}$ & 6.52$\times 10^{-14}$ & 5\\
8.61 &   4.25$\times 10^{-13}$ &  1.0$\times 10^{-15}$ &  6\\
11.6 &   2.11$\times 10^{-13}$ &  2.59$\times 10^{-15}$ &  5\\
11.6 &   3.08$\times 10^{-13}$ &  1.0$\times 10^{-14}$ &  7\\
18.4 &   2.76$\times 10^{-13}$ &  9.78$\times 10^{-15}$  &  6\\ 
22.1 &   4.21$\times 10^{-13}$ &  5.4$\times 10^{-15}$  &  5 \\
23.9 &   5.13$\times 10^{-13}$ &  3.0$\times 10^{-15}$  &  7\\ 
61.8 &   1.20$\times 10^{-12}$ &  1.0$\times 10^{-13}$  &  7 \\
70.0 &   1.04$\times 10^{-12}$ &  4.28$\times 10^{-15}$  &  8 \\
102. &   6.20$\times 10^{-13}$ &  1.0$\times 10^{-13}$  &  7 \\
160. &   3.07$\times 10^{-13}$ &  1.87$\times 10^{-15}$  &  8 \\
250. &   8.83$\times 10^{-14}$ &  3.60$\times 10^{-16}$  &  8 \\
350. &   3.30$\times 10^{-14}$ &  1.71$\times 10^{-16}$  &  8 
\enddata
\tablecomments{(1) \citealt{Myers2015}, (2) \citealt{Lasker2008}, (3) \citealt{GAIADR2}, (4) \citealt{Cutri2003}, (5) \citealt{Cutri2012}, (6) \citealt{Ishihara2010}, (7) \citealt{Hindsley1994}, (8) \citealt{Konyves2020}}
\end{deluxetable}

\begin{deluxetable}{ccc}
\tabletypesize{\footnotesize}
\tablecolumns{3} 
\tablecaption{Model Parameters\label{partab}} 
\tablehead{ 
\colhead{Parameter}    & \colhead{Value} & \colhead{Reference}}
\startdata 
Spectral Type & A7V & 1\\
T$_{eff}$ & 7500 K & 1\\
Radius (R$_{star}$) & 3.5 R$_{\odot}$  & \nodata\\
Mass & 2.0 M$_{\odot}$ & 1\\
Distance & 342 pc & 2\\
Disk inclination & 65$^{\circ}$& \nodata\\
Disk mass & 0.11 M$_{\odot}$& \nodata\\
Disk accretion rate & 2.0$\times 10^{-8}$ M$_{\odot}$& ...\\
Dust Grain file & www006.par & 3\\
Inner Radius & 0.016 au &\nodata\\
Outer Radius & 250 au &\nodata\\
Gap R$_{in}$ - R$_{out}$& 1.0 - 140 au&\nodata\\
Radial density $\propto R^{-A}$& A=1.0&\nodata\\
Scale height $\propto R^{B}$ & B=1.23&\nodata\\
Scale height (at R=R$_{\star})$ & 0.013 R$_{\star}$ &\nodata\\
Gap density ratio at R$_{out}$& 10$^{-5}$  &\nodata\\ 
Scale for radial exponential density cutoff & 30 au &\nodata\\
Number of radial grid cells & 400&\nodata\\
Number of theta (polar) grid cells & 197 &\nodata\\
Number of phi (azimuthal) grid cells & 2 &\nodata
\enddata
\tablecomments{(1) \cite{Ripepi2003}, (2) \cite{GAIADR2}, (3) Model 2 from \cite{wood02}. } 
\end{deluxetable}

\end{document}